\documentclass[twocolumn,showpacs,preprintnumbers,amsmath,amssymb]{revtex4}
\usepackage{graphicx}% Include figure files
\usepackage{dcolumn}% Align table columns on decimal point
\usepackage{bm}% bold math

\begin{document}
\title{Destruction of Superconductivity by Impurities
in the Attractive Hubbard Model} 
\author{Daniel Hurt, Evan Odabashian, Warren Pickett, Richard Scalettar}
\affiliation{Physics Department,
University of California,
Davis CA 95616, USA}
\author{Felipe Mondaini, Thereza Paiva, Raimundo R. dos Santos}
\affiliation{
Instituto de F\' \i sica,
Universidade Federal do Rio de Janeiro,
Cx.P.\ 68.528,
21945-970 Rio de Janeiro RJ, Brazil
}

\date{\today}

% submitted to PRB with PACS: 
% 74.20.-z  Theories and models of superconducting state
% 74.81.-g   Inhomogeneous superconductors and superconducting systems

\begin{abstract}
We study the effect of $U=0$ impurities on the superconducting and 
thermodynamic properties of the attractive Hubbard model on a square lattice.
Removal of the interaction on a critical fraction of 
$f_{\rm crit} \approx 0.30$ of
the sites results in the
destruction of off-diagonal long range order in the ground state.  
This critical fraction is roughly independent of filling in the range
$0.75 < \rho < 1.00$, although our data suggest that
$f_{\rm crit}$ might be somewhat larger below half-filling than
at $\rho=1$.
We also find that the two peak structure in the
specific heat is present at $f$ both below and above the
value which destroys long range pairing order.
It is expected that the high $T$ peak associated with local pair 
formation should be robust, but
apparently local pairing fluctuations are sufficient to generate 
a low temperature peak.
\end{abstract} 

\maketitle

The study of the 
interplay of disorder and interactions is one of the central problems in
condensed matter physics, and has been analyzed through a wide range of 
techniques.\cite{belitz94,sachdev89,ma81,yi94,logan93,zimanyi90,sandvik94}
Quantum Monte Carlo (QMC) is one approach which has been applied
only relatively recently, but now a number of investigations
of the disordered, repulsive Hubbard model
exist, both on finite lattices\cite{ulmke97,ulmke98,ulmke99} and within 
dynamical mean field theory where the momentum
dependence of the self-energy is neglected.\cite{ulmke95}
Several interesting effects, including
the possibility of the enhancement of
$T_{{\rm Neel}}$ by site disorder\cite{ulmke95}, the
occurrence of a Mott transition away from half-filling,\cite{byczuk04}
and a change in the temperature dependence of the resistivity from
insulating to metallic behavior
when interactions are turned on,\cite{scalettar99}
have been observed.
These studies have also explored different types of disorder,
including randomness in the chemical potential, hopping, 
Zeeman fields, and on-site interaction.\cite{malvezzi02}
At half-filling, some of these
disorder types preserve particle-hole symmetry which
has an important effect both on the
behavior of the static (magnetic and charge)
correlations,\cite{ulmke97} and on the 
conductivity.\cite{denteneer01}

The work described above concerns the interplay of randomness and
{\it repulsive} interactions.
Superconductor-insulator transitions 
provide a motivation to explore 
models with an effective {\it attractive} electron-electron interaction,
since these novel transitions are widely studied experimentally, 
especially in two 
dimensions.\cite{hebard94,haviland89,dynes84,valles92}
However, QMC studies of models like the attractive Hubbard
Hamiltonian with disorder are less 
numerous than for the repulsive case.\cite{dossantos93,scalettar99}
Previous work focused on locating the critical site disorder
strength for the insulating transition and determining
the value of the conductivity and its possible universality,\cite{trivedi96}
and on the interplay between the loss of phase coherence and
the closing of the gap in the density of states.\cite{huscroft98,ghosal02}

Another particularly interesting application of the attractive Hubbard 
model with disorder is the question of how inhomogeneities and 
pairing affect each other in high-temperature superconductors.
The attractive Hubbard model is the simplest Hamiltonian where
this interplay can be studied qualitatively.

Rather than looking at random chemical potentials,
as in the previous work just described,
here we will instead explore the effect of disordered interaction,
specifically turning off the attraction on a 
fraction $f$ of the sites in the $-U$ Hubbard model.\cite{litak}
Such vacancies are analogous to non-magnetic impurities in the repulsive
Hubbard model, and, indeed, at half-filling, 
the two problems are related by a particle-hole
transformation.  We determine
the filling dependence of the critical impurity 
concentration for the destruction
of superconductivity, 
through measurements of the equal time pair correlations and
the current-current correlations.  We also study the effects of
the disappearance of superconductivity on the specific heat.
Without disorder, a high temperature peak, associated with pair formation 
and a low temperature one, associated with pair coherence, are present. 
With disorder, we show that the high temperature 
peak is robust
as superconductivity disappears, and the low temperature one is weakened.

\section{The Model and Computational Approach}

The attractive Hubbard Hamiltonian we study is,
\begin{eqnarray}
H &=& -t\sum_{\langle {\bf rr'} \rangle \sigma}
( c_{{\bf r} \sigma}^{\dagger}c_{{\bf r'} \sigma}
+ c_{{\bf r'} \sigma}^{\dagger}c_{{\bf r} \sigma} ) 
-\mu \sum_{{\bf r} \sigma} n_{{\bf r} \sigma}
\nonumber\\
&-& \sum_{{\bf r}} \, U({\bf r}) \, (n_{{\bf r} \uparrow}-\frac12)
                    (n_{{\bf r} \downarrow}-\frac12)\  \cdot
\nonumber
\end{eqnarray}
Here $c^{\dagger}_{{\bf r}\sigma} 
(c_{{\bf r}\sigma})$ 
are fermion creation(destruction) operators at site ${\bf r}$
with spin $\sigma$, and $n_{{\bf r}\sigma}=
c_{{\bf r}\sigma}^{\dagger}c_{{\bf r}\sigma}$.
We use the coefficient $t$ of the kinetic energy term
to set our energy scale (i.e.~$t=1$).
The kinetic energy lattice sum $\langle {\bf rr'} \rangle$ is over
nearest neighbor sites on a two dimensional square lattice,
and $\mu$ is the chemical potential.
The on-site attraction $U({\bf r})$ is chosen to take on the two values
$U({\bf r})=0$ and
$U({\bf r})=-|U|$ with probabilities $f$ and $1-f$ respectively.\cite{foot1}
In this paper, we focus on a single interaction strength $|U|=4$.
Notice that we have chosen a particle-hole symmetric form 
of the interaction.  As a consequence, the energy level of a 
singly occupied vacancy site is higher by $|U|/2$ than a singly occupied
site with nonzero interaction.
We will report on results for the non-symmetric form
elsewhere.

We begin by briefly reviewing the physics of the pure attractive 
Hubbard model, $f=0$.
At half-filling, charge density wave and superconducting correlations
are degenerate, and this symmetry of the order parameters drives
the superconducting (and CDW) transition temperatures
to zero.  Doping breaks the symmetry in favor of 
pairing correlations, and there is now a Kosterlitz-Thouless
transition to a superconducting state at finite temperature.
It is well established that $T_c$ rises rapidly with doping
away from half-filling, reaching
a maximal value at $\rho = 1 \pm \delta$ with $\delta \approx 1/8$.
$T_c$ decreases gradually thereafter.\cite{scalettar89,moreo91}
There is some debate as to the exact value of the maximal
$T_c$, with estimates\cite{paiva04} 
in the range $0.05 t < T_{c,{\rm max}} < 0.2t$.
The problem lies with the rather large system sizes needed
to study Kosterlitz-Thouless transitions numerically,
and hence to benchmark the accuracy of the approximate
analytic calculations.
Our focus here will not be on this finite temperature 
phase transition, but rather on the quantum phase transition 
out of the superconducting ground state which
occurs through the introduction of 
a non--zero fraction $f$ of vacancies.

To distinguish a superconducting phase, we first look for long range
structure in the equal time pair-pair correlation function,
\begin{eqnarray}
c({\bf r}) &=& \langle \, \Delta({\bf r}) \Delta^{\dagger} (0) \, \rangle,
\nonumber\\
\Delta^{\dagger}({\bf r}) 
&=& c_{{\bf r}\uparrow}^{\dagger} c_{{\bf r}\downarrow}^{\dagger}.
\nonumber
\end{eqnarray}
This pair-pair correlation function can also be
summed spatially to define the structure factor,
\begin{eqnarray}
P_s = \sum_{{\bf r}} c({\bf r})
= \sum_{{\bf r}} \langle \, \Delta({\bf r}) \Delta^{\dagger} (0) \, \rangle.
\nonumber
\end{eqnarray}
The finite size scaling of $P_s$ is described in the following section.

The current-current correlations probe the superfluid weight
and provide an alternate means to detect the
destruction of superconductivity.\cite{scalapino93}  We define,
\begin{eqnarray}
{\Lambda}_{xx}({\bf r}, \tau)
&=& \langle j_{x} ({\bf r}, \tau) j_{x} (0, 0) \rangle
\nonumber\\
j_{x}({\bf r} \, \tau) &=& e^{H \tau}
\left[it \sum_\sigma
(c_{{\bf r}+\hat x,\sigma}^{\dagger}c_{{\bf r},\sigma}^{\ } -
c_{{\bf r},\sigma}^{\dagger} c_{{\bf r}+\hat x,\sigma}^{\ } )
\right] e^{-H \tau}
\nonumber
\end{eqnarray}
and the Fourier transform in space and imaginary time,
\begin{eqnarray}
\Lambda_{xx}({\bf q},\omega_n) = \sum_{\bf r} \int_0^\beta d \tau
e^{i {\bf q}\cdot{\bf r} }
e^{-i \omega_n \tau}
\Lambda_{xx}({\bf r},\tau),
\nonumber
\end{eqnarray}
where $\omega_n = 2n\pi/\beta$.

The longitudinal part of the current-current correlation function
satisfies the f-sum rule, which relates its value to the kinetic energy
$K_x$, 
% for all temperatures, interaction strengths, and lattice sizes:
\begin{eqnarray}
\Lambda^{{\rm L}} & \equiv &
{\rm lim}_{q_x \rightarrow 0} \hskip0.1in
{\Lambda}_{xx} (q_{x},q_{y}=0,\omega_n = 0) \nonumber\\
\Lambda^{\rm L} &=&
K_{x}.
\nonumber
\end{eqnarray}
Here $K_{x} = \langle -t \sum_\sigma
(c_{{\bf r}+\hat x,\sigma}^{\dagger}c_{{\bf r},\sigma}^{\ } +
c_{{\bf r},\sigma}^{\dagger} c_{{\bf r}+\hat x,\sigma}^{\ } ) \rangle$,
Meanwhile, in the superconducting state the transverse part,
\begin{equation}
\Lambda^{T}\equiv
{\rm lim}_{q_y \rightarrow 0} \hskip0.1in
{\Lambda}_{xx} (q_{x}=0,q_{y},\omega_n=0) \  \cdot
\nonumber
\end{equation}
can differ from the longitudinal part,
the difference being
the superfluid stiffness $D_s$,
\begin{eqnarray}
D_{s}/\pi & = & [
\Lambda^{\rm L}-
\Lambda^{\rm T}] \nonumber \\
& =  &
[K_{x}-
\Lambda^{\rm T}]\  .
\nonumber
\end{eqnarray}
Thus the current-current correlations provide an alternative, 
complementary method to the equal time
pair correlations for looking at the superconducting transition.

In order to evaluate these quantities, we use the determinant QMC
method.\cite{blankenbecler81}  In this approach, we discretize the
inverse temperature $\beta = L \Delta \tau$, and
a path integral expression is
written down for the partition function $Z$.  The electron-electron
interactions are decoupled by the introduction of a 
Hubbard-Stratonovich field.  The fermion
degrees of freedom can then be integrated out analytically,
leaving an expression for $Z$ which involves an integral over
the Hubbard-Stratonovich field, 
with an integrand which is the product 
of two determinants of matrices of dimension the system size.
We perform the integral stochastically.  
In the case of the attractive Hubbard model
considered here, the traces over the spin up and spin down electrons are
given by the determinant of the same matrix,
the integrand is a perfect square,
and hence there is no sign problem.

Observables are evaluated
as the appropriate combinations of Greens functions which are given
by matrix elements of the inverse of the matrix appearing
as the integrand.  Systematic errors in the pairing and
current measurements associated with the discretization of $\beta$,
with our choice of $\Delta \tau$, are typically 
smaller than the error bars associated with the statistical fluctuations
for a single disorder realization and the error bars associated
with sample-to-sample variations.  The `Trotter errors'
are discernable in the energy, but we have verified they do not
affect the qualitative behavior of the specific heat.
Our procedure for measuring the specific heat is described in 
a subsequent section.

As remarked above,
there is no `sign problem' for this attractive Hubbard
model, so we can do computations at arbitrarily low temperature 
(large $\beta$).  In practice, we find that for the parameter values
and lattice sizes under consideration here the superconducting correlations
reach their asymptotic ($T=0$) values when $T < 1/12$ ($\beta > 12$),
though we collect data also at $T=1/16$ to check this conclusion.
These temperatures are about one hundredth of the bandwidth
$W=8t$.

% We comment on our choice of chemical potential.
% For the correlation
% functions, we are mainly concerned with large values of $\beta$, in which 
% case $\rho(\mu)$ is hardly dependent on the particular disorder 
% realization. To this end, the chemical potential is chosen so 
% that the averaged densities were the desired ones. 
% This is not the case for the specific heat since it is 
% calculated over a much wider range of temperatures, demanding a more 
% judicious choice of $\mu$.

\section{Equal Time Pairing Correlations}

\begin{figure}
\includegraphics[width=3.0in,height=3.6in,angle=-90]{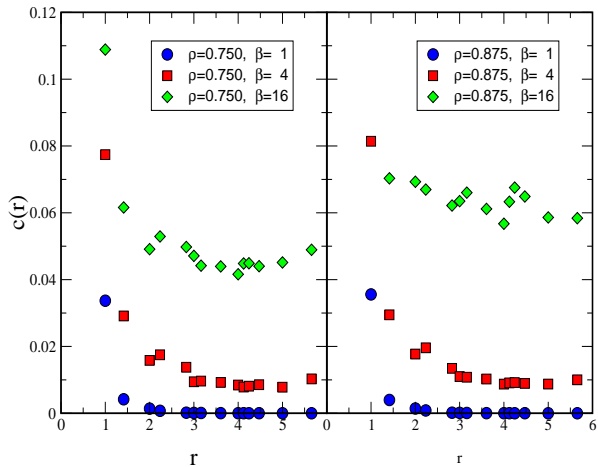}
\caption{
The equal time pair correlation $c({\bf r})$ is shown as 
a function of the separation $|{\bf r}|$ of the points of injection
and removal of the pair.  Here the lattice size is $N$=8x8, $f=1/16$, and
the densities $\rho=0.750$ (left) and $\rho=0.875$ (right).  
Long range correlations
build up as the temperature is lowered, even though the attractive
interactions have been somewhat diluted.
Error bars are the size of the symbols.  Scatter in the data
is associated with the anisotropy in the correlations, which are
not a function of the distance only.
}
\label{fig1}
\end{figure}

\begin{figure}
\includegraphics[width=3.0in,height=3.6in,angle=-90]{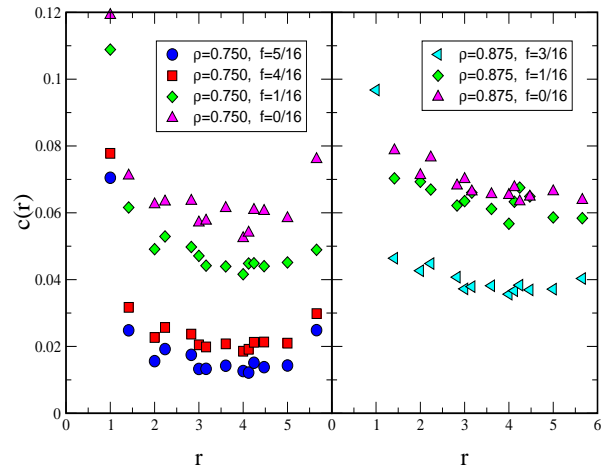}
\caption{
The equal time pair correlation $c({\bf r})$ is shown as 
a function of the separation $|{\bf r}|$ of the points of injection
and removal of the pair.  Here the lattice size is $N$=8x8,
the density $\rho=0.750$ (left), and the temperature $T=1/16$.
Long range correlations are decreased with increasing vacancy
fraction $f$.  Density $\rho=0.875$ is at right.
}
\label{fig2}
\end{figure}

We first show
the spatial behavior of the pairing correlations $c({\bf r})$.
In Fig.~\ref{fig1}, the vacancy fraction 
$f=1/16$ is fixed, and we study how long range correlations in 
$c({\bf r})$ develop as the temperature $T$ is reduced.
Results for two different densities $\rho=0.750$ and $\rho=0.875$ are given.
The lattice size $N$=8x8.
% and we have averaged over XX different disorder
%realizations for each $\beta$.  
In Fig.~\ref{fig2} we fix $T=1/16$ and examine
$c({\bf r})$ for different $f$.  As the vacancy rate increases,
the long range order vanishes. 
% The chemical potential was chosen so 
% that the averaged densities were the desired ones. For the correlation
% functions, we are mainly concerned with large values of $\beta$, in which 
% case $\rho(\mu)$ is hardly dependent on the particular disorder 
% realization. This is not the case for the specific heat since it is 
% calculated over a much wider range of temperatures, demanding a more 
% judicious choice of $\mu$.

\begin{figure}
 \includegraphics[width=2.4in,height=3.6in,angle=-90]{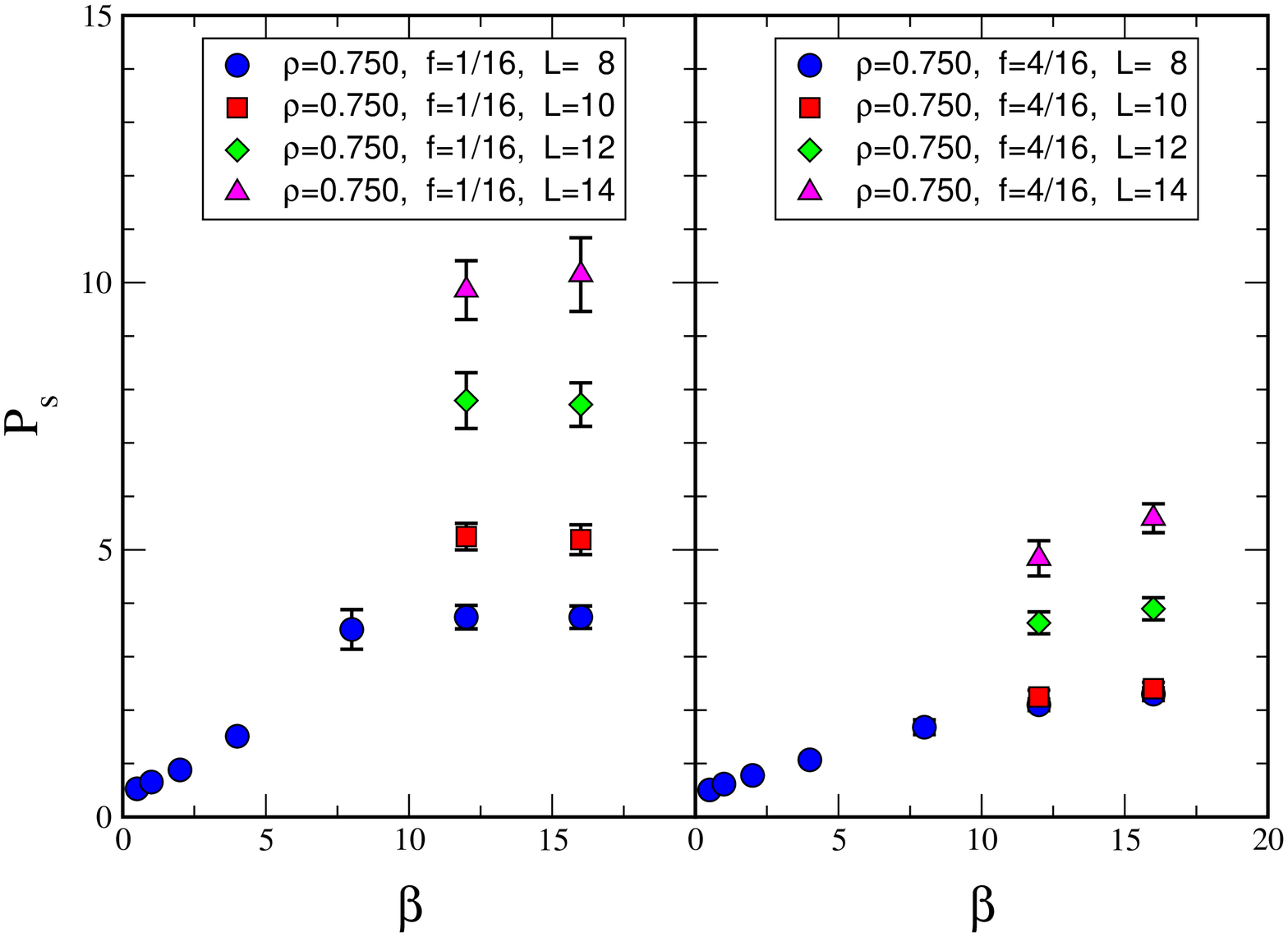}
% \vskip-0.3in
 \includegraphics[width=2.4in,height=3.6in,angle=-90]{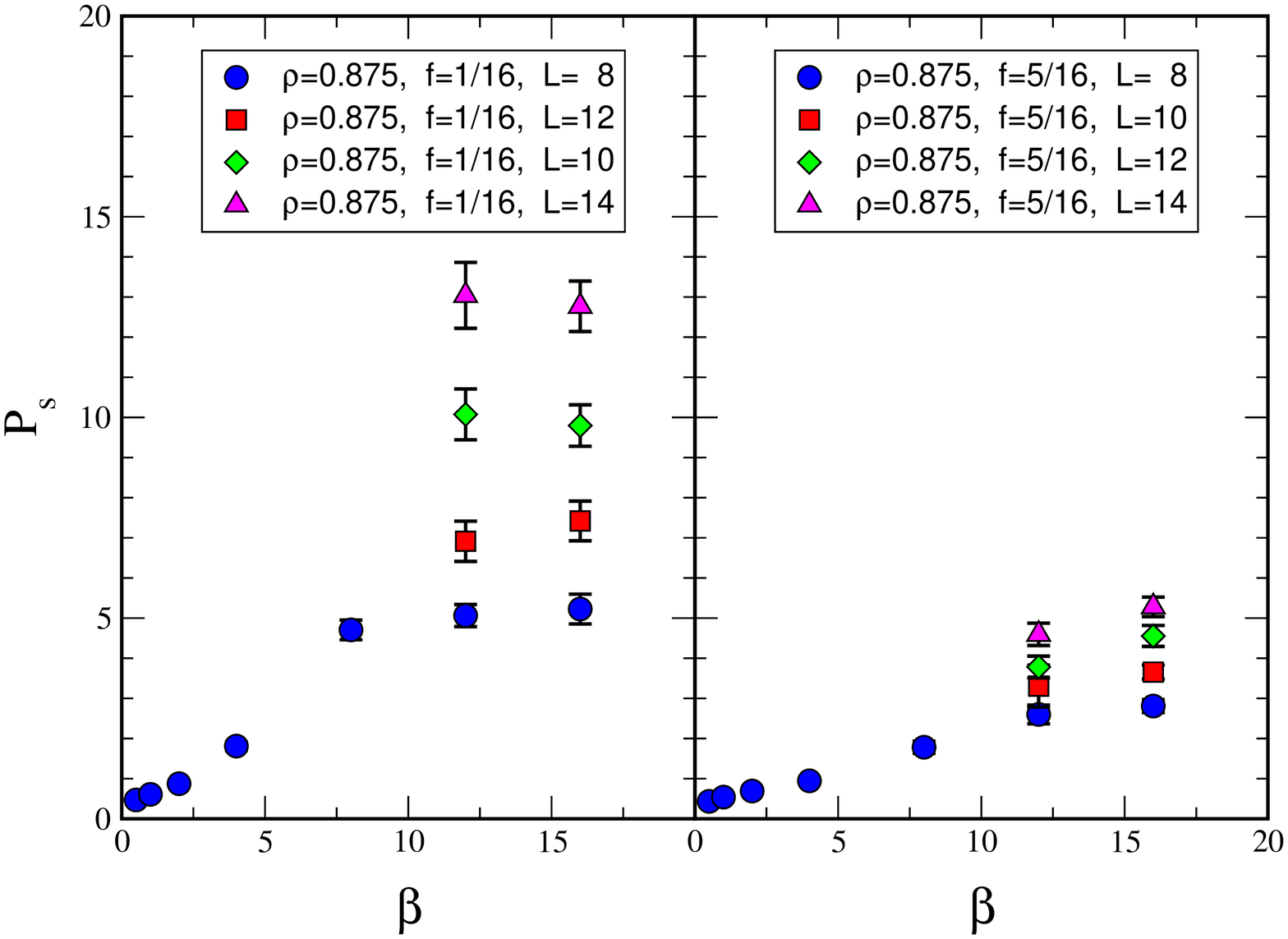}
\caption{
(a)  The equal time pair structure factor $P_s$ is shown as 
a function of inverse temperature $\beta$.
Here the density $\rho=0.750$, and dilution
fraction $f=1/16$.
$P_s$ grows significantly with lattice size $N$ at low temperatures.
The temperature at which data for the different lattice sizes separate
indicates the point at which the coherence length $\xi(T)$ becomes
comparable to the linear lattice size.
(b)  $\rho=0.875$, $f=1/16$.
(c)  $\rho=0.750$, $f=4/16$.
(d)  $\rho=0.875$, $f=5/16$.
}
\label{fig3}
\end{figure}

These results can be analyzed more carefully by performing
a finite size scaling study of the pair structure factor $P_s$.
If the pair correlations $c({\bf r})$ extend over the entire lattice, $P_s$,
which sums this quantity,
will grow linearly with size.  On the other hand,
if the pair correlations are finite
in range, $P_s$ is independent of size.
Fig.~\ref{fig3} shows $P_s$ as a function of $\beta$ for different
lattice sizes.
In Fig.~\ref{fig3} we choose a small vacancy fraction 
$f=1/16$ (left panels) and see that at large $\beta$ the pair 
structure factor increase significantly with lattice size.
In the right panels of Fig.~\ref{fig3}, at large vacancy fractions,
$f=4/16$ and $f=5/16$, the pair structure factor grows much less rapidly
with lattice size at low temperatures.

\begin{figure}
\includegraphics[width=4.2in,height=3.8in,angle=-90]{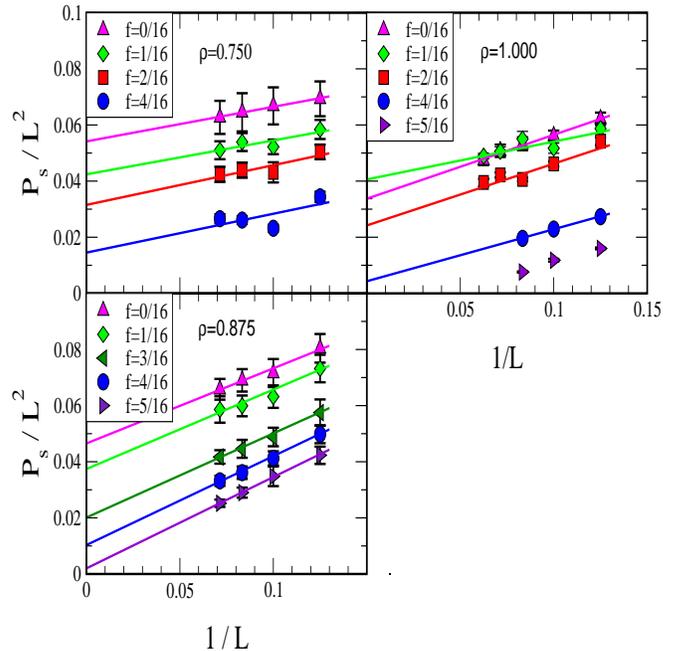}
\caption{
These finite size scaling plots show the equal time pair 
structure factor $P_s$ at low temperature $T=1/16$
as a function of linear lattice size
$1/L=1/\sqrt{N}$.
Different curves are for different values of the vacancy fraction.
For small $f$, $P_s$ extrapolates to a nonzero value in the thermodynamic
limit $1/L \rightarrow 0$.  For larger $f$ there is no longer
a nonzero extrapolated value.
The curves are upper left panel $\rho=0.750$,
lower left panel $\rho=0.875$, and
upper right panel $\rho=1.000$.
An interesting  feature of the half-filling data
is the corssing of the $f=0$ and $f=1/16$ curves,
indicating an initial enhancement of pairing by the impurity sites.
}
\label{fig4}
\end{figure}

Fig.~\ref{fig4} determines the critical vacancy fraction through a finite
size scaling analysis.  As shown by Huse,\cite{huse88} the spin-wave
correction to the pair structure factor is expected to be proportional
to the linear lattice size:
\begin{equation}
\frac{P_s}{L^2}=\Delta_0^2+ \frac{a}{L},
\end{equation}
where $\Delta_0$ is the superconducting gap function  at zero temperature, 
and $a(U,f)$ is independent of $L$.
In Fig.~\ref{fig4}
we plot the low temperature value of
$P_s$ versus $1/\sqrt{N}$ for lattice sizes ranging from
$N$=8x8 to $N$=14x14, and
$\rho=0.750$, $\rho=0.875$, and $\rho=1.000$.
At small $f$, there is the expected
nearly linear behavior, extrapolating to a nonzero value
in the thermodynamic limit.\cite{foot2}
As $f$ increases, the nonzero extrapolation disappears.
This figure contains one of the central results of this paper, namely 
the determination of the critical values of vacancy fraction $f$
for the destruction of superconductivity.
To within the resolution of these simulations,
the critical dilution fraction $f_{\rm crit} \approx 0.30$ for the
destruction of superconductivity in the ground state is the
same for fillings $\rho=0.750$ and $\rho=0.875$.  

Half-filling appears to behave somewhat differently.
Disorder first enhances superconductivity, as seen
by the crossing of the $f=0$ and $f=1/16$ lines in Fig.~4,
before driving it to zero a bit sooner than for the doped cases.
Impurity sites break the special
superconducting and charge density wave degeneracy at $\rho=1$,
thereby providing a possible reason for the initial enhancement.

It should be noted that the values of $f_{\rm crit}$ 
indicate that the transition is
neither of the classical site-percolation type, for which $f_c^{\rm
classical,site}=0.41$, nor it is related to the quantum percolation one,
for which $f_c^{\rm quantum,site}=f_c^{\rm quantum,bond}=0$, the latter
reflecting the fact that any disorder localizes tight-binding electrons.
\cite{soukoulis}

\section{Current Correlations and Superfluid Weight}

\begin{figure}
\includegraphics[width=4.0in,height=3.6in,angle=-90]{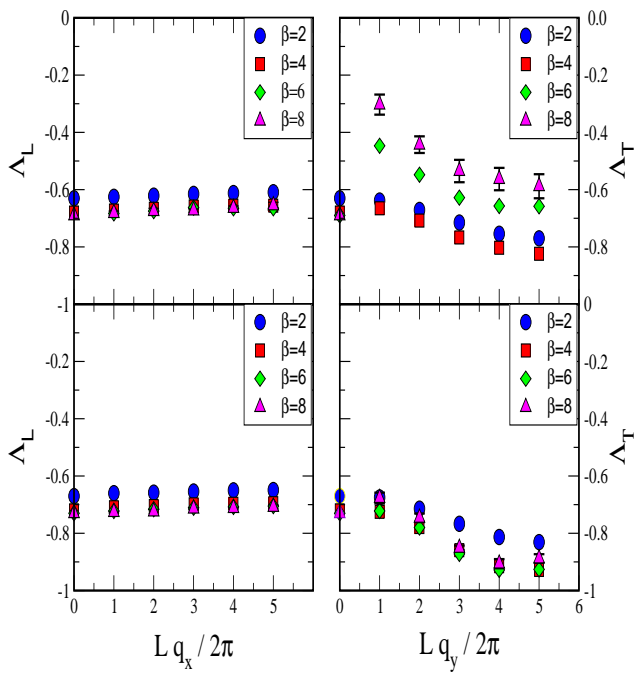}
\caption{Left panels:
Longitudinal current-current correlations
$\Lambda_{xx}(q_x,q_y=0,i\omega_n=0)$ 
as a function of $q_x$.
Here the density $\rho=0.875$. 
$\Lambda_{xx}(q_x,q_y=0,i\omega_n=0)$ 
extrapolates to the kinetic energy (symbols at $q=0$)
for all temperatures and both fillings when $f=0.04$ (top), 
which is superconducting,
and $f=0.32$ (bottom), which is not.
Right panels:
Transverse current-current correlations
$\Lambda_{xx}(q_x=0,q_y,i\omega_n=0)$ 
extrapolate to the kinetic energy (solid square at $q_x=0$)
at high temperature for $f=0.04$ (top),
but break away at low $T$ indicating the presence of a 
non-zero superfluid density.
For $f=0.32$ (bottom)
the dilution fraction exceeds $f_{\rm crit}$
so now at low temperature there is no superfluid density.
All data are for 10x10 lattices.
}
\label{fig5}
\end{figure}

These conclusions are confirmed by studying
the current-current correlations.
We first show, in the left panels of
Fig.~\ref{fig5}, that the longitudinal current-current
correlation $\Lambda_{xx}(q_x,q_y=0,i\omega_n=0)$ extrapolates
to the kinetic energy as $q_x \rightarrow 0$.
This sum rule is satisfied in all parameter regimes,
both small and large vacancy rate $f$,
and both small and large temperatures $T$.

In contrast, the transverse response 
$\Lambda_{xx}(q_x=0,q_y,i\omega_n=0)$,
given in the right hand panels of Fig.~\ref{fig5}, extrapolates
to the kinetic energy with $q_y$ 
only when the temperature is high at small $f$.
As pairing correlations develop across the lattice with
decreasing $T$ (Fig.~\ref{fig1})
the transverse response breaks away from the kinetic energy,
indicating a nonzero superfluid weight $D_s$.
At vacancy concentrations beyond the point at which superconductivity 
is destroyed, $D_s$ remains zero with decreasing temperature.

\begin{figure}
\includegraphics[width=2.4in,height=3.0in,angle=-90]{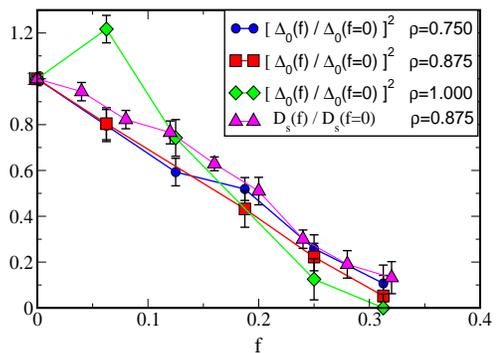}
\caption{
Values for the $L\rightarrow \infty$ pair structure factor versus
vacancy fraction $f$
at density $\rho=0.750$ (circles), $\rho=0.875$ (squares)
and $\rho=1.000$ (diamonds).  The inverse temperature
$\beta=10$. 
Also shown is the superfluid fraction $D_s$ on
a fixed 10x10 lattice size.
Both measurements give a consistent estimate of $f_c \approx 0.30$,
}
\label{fig6}
\end{figure}

In Fig.~\ref{fig6} we show
the finite-size extrapolated values of $P_s$ inferred from Fig.~\ref{fig4}.
These measurements
give an indication of the location of $f_{\rm crit}$
in the thermodynamic limit.
To within our numerical uncertainties,
$f_{\rm crit}$ is the same for the fillings $\rho=0.750$
and $\rho=0.875$, and may be a bit less
for $\rho=1.000$. 
% Another aspect which distinguishes the doped case
% from half-filling is that the data seem to lie on a straight line in the 
% former, while there is a significant curvature in the latter.
In Fig.~\ref{fig6}, for density $\rho=0.875$, we also show
the low temperature values of the superfluid
fraction $D_s$ obtained from Fig.~\ref{fig5}.
The two measurements, $\Delta_0$ and
$D_s$, give consistent results for $f_{\rm crit}$.

\section{Specific Heat}

At half-filling,
the specific heat of the repulsive
(attractive) Hubbard Hamiltonians, in the absence of impurities,
exhibit two features.\cite{takahashi74,duffy97,staudt00}
There is a peak associated
with moment(pair) formation at high temperatures, $T \approx U/3$.
At lower temperatures, $T \approx J/4 = t^2/U$, there is a second
peak associated with magnetic ordering (pair coherence).
Interestingly, in two dimensions,
this two peak structure appears to survive even down
to weak coupling where the two energy scales merge,\cite{paiva04}
in contrast to the behavior in one 
dimension,\cite{shiba72,schulte96,usuki89,koma90,sanchez98}
and in the paramagnetic phase in infinite 
dimensions.\cite{georges93,vollhardt97,chandra99}

What happens to this behavior when disorder is introduced, specifically
when $U=0$ sites are inserted, as in our present model?  
To address this question we compute the specific heat by fitting
QMC data for $E_{\rm qmc}(T_n)$ to the functional form,
\begin{equation}
E_{\rm fit}(T)=E_{\rm fit}(0)+\sum_{l=1}^{M} c_{l} e^{-\beta l \Delta} \, .
\nonumber
\end{equation}
The parameters  $\Delta$ and $c_l$
are selected to minimize
\begin{equation}
\chi^2 = \frac{1} {N_T} \sum_{n=1}^{N_T} { \frac{(E_{\rm fit}(T_n) - 
E_{\rm qmc}(T_n))^2}{
(\delta E_{\rm qmc}(T_n))^2} } \,.
\nonumber
\end{equation}
We choose a number of parameters $M$ equal to about one-fourth of the
QMC data points to allow a good fit to the data without overfitting.
We check that different $M$ around this value all provide comparable results.\cite{foot3}

\begin{figure}
\includegraphics[width=2.4in,height=2.4in,angle=-90]{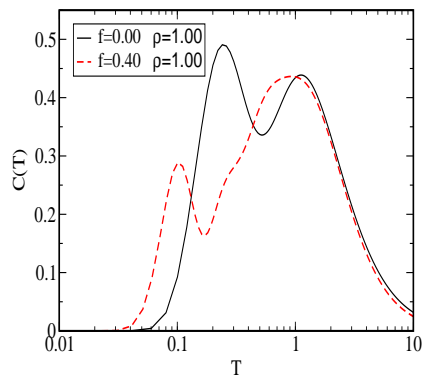}
\caption{
Specific heat of the attractive Hubbard model
on $N$=10x10 lattices and density $\rho=1.000$ 
for $f=0.00$ (solid curve) and $f=0.40$ (dashed curve).
The high temperature peak indicates the temperature of
local pair formation and is, as expected, not very much affected by
the dilution of the attractive interaction on some of the sites.
The low $T$ peak associated
with pair coherence is substantially reduced
by dilution, but appears to be present even at large $f$,
where long range order has been lost.
}
\label{fig7}
\end{figure}

Fig.~\ref{fig7} summarizes our results for the specific heat.  
At a vacancy fraction $f=0.40$, the high temperature peak in 
the specific heat, which is associated with the formation of local pairs, 
is, as expected, somewhat broadened relative 
to the clean system ($f=0.00)$, but remains otherwise robust.
The low temperature peak, associated with pair coherence,
is much more substantially reduced, and shifted to lower $T$.  
However, it appears to remain present even for this value of 
$f$ which is beyond $f_{\rm crit}$.
% We conclude that only local pair fluctuations are required
% for the formation of this low temperature peak.

% One of the interesting aspects of the 
% infinite dimension studies is a universal crossing
% of the specific heat curves for different 
% interaction strengths.\cite{georges93,vollhardt97,chandra99}
% We see in Fig.~\ref{fig8} that this universal crossing is evident in the
% disordered model as well, although here the parameter being tuned is
% $f$ rather than $U$.  Vollhardt has discussed the
% physics behind this crossing, which, loosely speaking,
% may be attributed to the sum rule that the integrated specific heat
% gives the same high temperature entropy, ln4 per site, for all parameter
% values.  This effect can be seen in Fig.~7 where the reduction in
% the low $T$ peak in the specific heat is accompanied by an increase in 
% the height of the high $T$ peak.
% The universality of the
% crossing is a more subtle issue.\cite{vollhardt97}

% \begin{figure}
% \includegraphics[width=2.0in,height=2.0in,angle=-90]{fig8.ps}
% \caption{
% As for the case of changing interaction strengths in
% the pure Hubbard model, the specific heat of the Hubbard model
% with disordered vacancy sites studied here exhibits a universal crossing
% of the specific heat.  Parameters are the same as for Fig.\ref{fig7}.
% }
% \label{fig8}
% \end{figure}

\section{Conclusions}

In  this paper we have looked at the destruction of the superconducting state
in the attractive Hubbard model through the systematic inclusion
of $U=0$ sites.  A consistent picture emerges from
an analysis of the pairing structure factor and the superfluid
fraction, namely that
pairing survives out to a dilution of about $3/10$ of the attractive
sites, and that this value is not very dependent on the filling, although
it may be somewhat increased away from half-filling.
As the fraction $f$ of $U=0$ sites is increased
the high temperature peak in $C(T)$, which signals local pair formation,
remains relatively unchanged.  The low $T$ peak which signals
the establishment of pairing order  gets pushed down.
Although the inclusion of non-interacting sites might remind
one of percolation, the transition here is not percolative.
The critical value of $f$ is neither that of classical site-percolation  
nor of quantum percolation.
Finally, we have shown that directly at half-filling, 
pairing correlations
are enhanced by the inclusion of vacancy sites.  We attribute this
effect to the breaking of degeneracy of the CDW and
superconducting order in favor of pairing.

\vskip0.2in
\noindent
\underbar{Acknowledgments:}
We acknowledge useful conversations with D. Leitsch.  This work was
supported by NSF-DMR-0312261, NSF-INT-0203837, NSF-DMR-0421810
and by the Brazilian Agencies, CNPq, FAPERJ, Instituto de 
nanoci\^encias/MCT, and Funda\c c\~ao Universit\'aria Jos\'e 
Bonif\'acio/UFRJ.
Evan Odabashian was supported by the NSF REU program, NSF-PHY-0243904.


\begin{thebibliography}{10}
\bibitem{foot1}
In order to study the physics at a particular lattice size $L$x$L$
and value of $f$,
we choose $f\,L^2$ sites randomly and set $U=0$ on those sites.  
We typically use 10--20 such realizations to average over the different
disorder configurations.
If $f \, L^2$ is not an integer, we average over the two adjacent
integer values, with appropriate weights.
\bibitem{foot3}
We note that other techniques might be (and have been) 
used, e.g.~fitting to polynomial forms\cite{duffy97},
numerical differentiation of the energy data, or computation of $C$
from energy fluctuations.
\end{thebibliography}
\end{document}